\documentclass[prl,aps,twocolumn,amsmath,amssymb,floatfix]{revtex4-2}
\usepackage{physics,graphicx,hyperref}
\hypersetup{colorlinks=true,citecolor=blue,linkcolor=blue,urlcolor=blue}
\usepackage[utf8]{inputenc}
\usepackage{mathtools}
\usepackage[normalem]{ulem}

\usepackage[cal=euler]{mathalfa} 

\begin{document}

\title{Type-II Mirror Chern Insulator in Altermagnets}

\author{Amrita Mukherjee}
\email{Contact author: amritaphy92@gmail.com}
\affiliation{Department of Condensed Matter Physics and Materials Science, Tata Institute of Fundamental Research, Mumbai 400005, India}

\author{Pritesh Srivastava}
\affiliation{Department of Condensed Matter Physics and Materials Science, Tata Institute of Fundamental Research, Mumbai 400005, India}

\author{Rahul Verma}
\affiliation{Department of Condensed Matter Physics and Materials Science, Tata Institute of Fundamental Research, Mumbai 400005, India}

\author{Bahadur Singh}
\email{Contact author: bahadur.singh@tifr.res.in}
\affiliation{Department of Condensed Matter Physics and Materials Science, Tata Institute of Fundamental Research, Mumbai 400005, India}

\title{Type-II Mirror Chern Insulator in Altermagnets}

\begin{abstract}
Altermagnets with momentum-dependent spin splitting despite zero net magnetization can support unique topological states under broken time-reversal symmetry. We predict a mirror-symmetry-protected topological crystalline insulator with momentum-separated edge modes in a two-dimensional altermagnet. Using a square-octagon lattice model, we show that altermagnetic order generates symmetry-related valley-polarized Dirac nodes, which are gapped by spin-orbit coupling to yield a mirror Chern insulator with $C_{\mathcal{M}}=2$. In contrast to conventional mirror Chern insulators, where the two mirror-protected edge modes cross at the same momentum to form a Dirac cone, altermagnetic spin splitting and valley-selective band inversion separate these edge modes in momentum. We refer to this phase as a type-II mirror Chern insulator. We further propose a PbSe/$\mathrm{V_2Se_2O}$ heterobilayer as a candidate material for realizing this phase through the altermagnetic proximity effect. Our results establish altermagnetism as a route to mirror-protected topological phases with momentum-separated edge modes.
\end{abstract}

\maketitle
\textit{Introduction.} Topological crystalline insulators (TCIs) are quantum phases where crystalline symmetries protect metallic boundary states beyond the paradigm of time-reversal symmetry ($\mathcal{T}$)~\cite{singh2023topology,fu2011topological,hsieh2012topological,xu2012observation,TCIsuper,niu2015two,khalaf2018,singh2019TCI}. Their topological properties originate from bulk band inversion and are characterized by topological invariants defined within crystal-symmetry subspaces. Through bulk-boundary correspondence, these invariants guarantee gapless states on symmetry-preserving boundaries. A prototypical example is SnTe, where a nonzero mirror Chern number defined on the $(110)$ mirror plane ($\mathcal{M}_{110}$) protects surface states on mirror-symmetric terminations~\cite{hsieh2012topological,xu2012observation}. Crystalline symmetry also enables magnetic TCIs, where nontrivial topology survives despite broken $\mathcal{T}$ symmetry, provided the protecting crystal symmetry is preserved~\cite{Zhang_MTCI,Chen_MTCI,Sarkar2022,xu2019higher,riberolles2021magnetic, li2019magnetically,hu2020realization}. In nonmagnetic mirror TCIs, the surface Dirac cone forms a Kramers pair satisfying $E_{\uparrow}(\mathbf{k})=E_{\downarrow}(-\mathbf{k})$ and becomes degenerate at time-reversal-invariant momenta. Along the mirror-invariant line, two edge modes carrying opposite mirror eigenvalues cross to form a mirror-protected Dirac cone. Although Kramers degeneracy is lifted in magnetic systems, the Dirac crossing remains protected by mirror symmetry, as demonstrated in EuIn$_2$As$_2$~\cite{Sarkar2022,xu2019higher}. Nevertheless, the two branches of the mirror-protected Dirac cone remain locked to the same band-inversion momentum.

\begin{figure}[t!]
\centering
\includegraphics[width=0.49\textwidth]{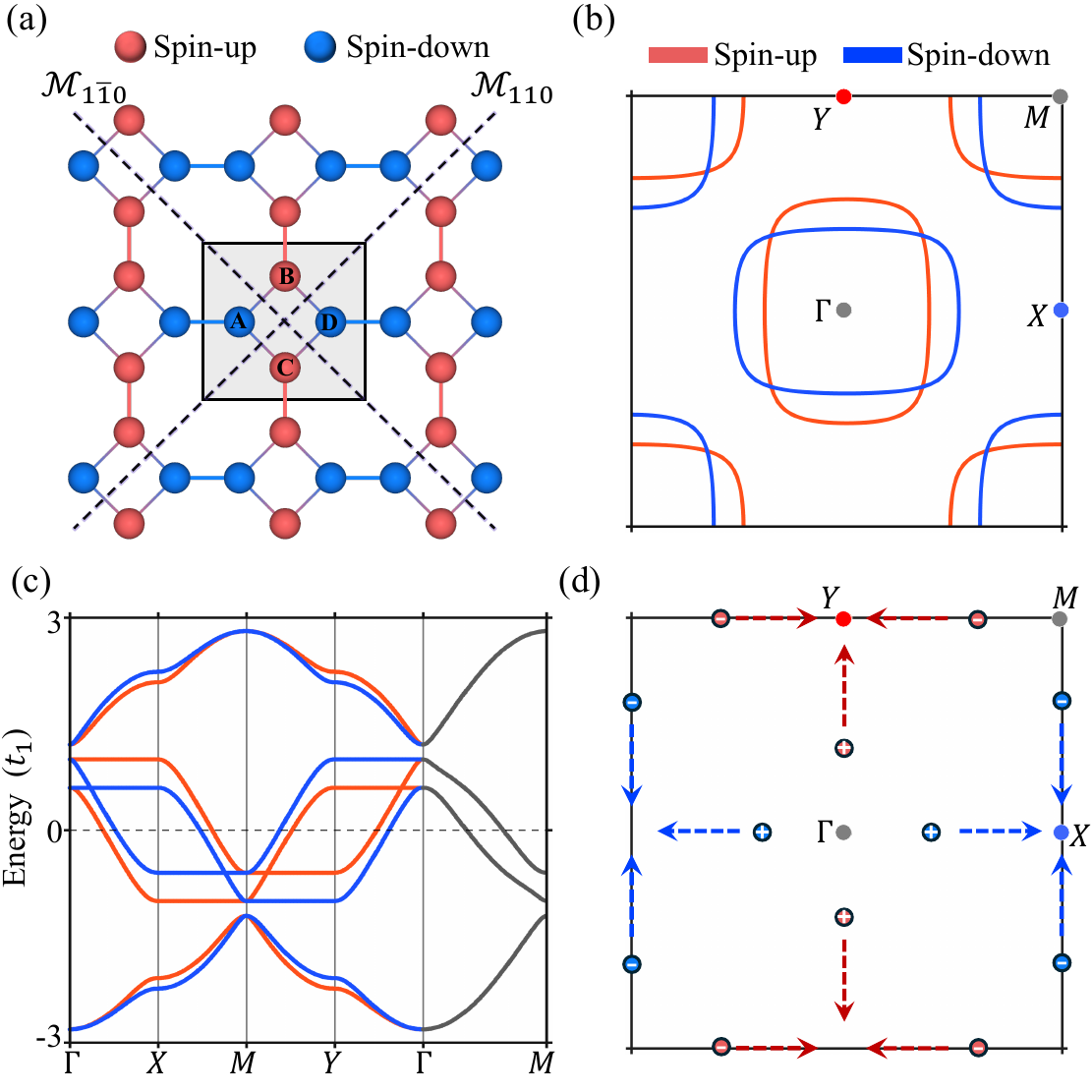}
\caption{(a) Square-octagon lattice with altermagnetic order. Red (blue) sites denote positive (negative) exchange fields. The blue and red bonds denote the intercell hopping ($t_2$) between the blue and red sites, respectively, while the color gradient within each square plaquette represents the intracell hopping ($t_1$). The black square outlines the unit cell, and the dashed lines indicate the mirror planes $\mathcal{M}_{110}$ and $\mathcal{M}_{1\bar{1}0}$. (b) Spin-resolved Fermi contours for $t_2=0.8$ and $\delta_m=0.2$. The high-symmetry points $\Gamma$ and $M$ are marked by gray circles, while the symmetry-related $X$ and $Y$ valleys are highlighted by blue and red circles, respectively. (c) Corresponding spin-resolved band structure along the high-symmetry path. (d) Evolution of the bulk Dirac nodes without SOC. The symbols $\pm1$ denote the winding number $w=\gamma/\pi$, and the dashed arrows indicate the evolution of the Dirac nodes with increasing $\delta_m$.}
\label{fig1}
\end{figure}

Recently discovered altermagnets constitute a new class of collinear magnetic materials that combines the compensated magnetic order of antiferromagnets with momentum-dependent spin splitting reminiscent of ferromagnets~\cite{vsmejkal2020crystal,smejkal2022beyond,vsmejkal2022emerging,roig2024,thomale2025,zhu2025,jungwirth2026symmetry}. This unique combination has opened new opportunities for spintronics, unconventional superconductivity, multiferroics, and topological quantum phases~\cite{vsmejkal2022emerging,antonenko2025mirror,type2qsh,Hu2026SVLNL,srivastava2026isolation,Jungwirth2026AltermagneticSpintronics}. The spin splitting originates from spin-lattice symmetry, which relates opposite-spin sublattices through crystal rotations or mirror operations and produces characteristic $d$-, $g$-, and $i$-wave spin textures~\cite{vsmejkal2022emerging}. In the $d$-wave phase considered here, opposite-spin states satisfy $E_{\uparrow}(\mathbf{k})=E_{\downarrow}(C\mathbf{k})$ for $C=(C_{4z},\mathcal{M}_{110},\mathcal{M}_{1\bar{1}0})$ instead of the Kramers relation $E_{\uparrow}(\mathbf{k})=E_{\downarrow}(-\mathbf{k})$. As a result, the spin splitting is finite over most of the Brillouin zone and vanishes only along symmetry-enforced directions, producing opposite spin polarizations at the $C_{4z}$-related $X$ and $Y$ valleys [Fig.~\ref{fig1}(c)]. These symmetry-related valleys also carry opposite Berry curvature~\cite{Mak2014VHS}. Altermagnetism thus intertwines spin, valley, and crystalline symmetry. Whether this interplay can generate unique mirror TCIs and reshape mirror-protected boundary states remains an open question.

In this work, we show that altermagnetic order and mirror symmetry stabilize a mirror TCIs with momentum-separated edge modes, which we identify as a type-II mirror Chern insulator. Using a two-dimensional (2D) square-octagon lattice with $\mathcal{M}_{001}$ mirror symmetry, we demonstrate that altermagnetic order and spin-orbit coupling (SOC) stabilizes a mirror Chern insulator with $C_{\mathcal{M}}=2$. Unlike conventional mirror Chern insulators, altermagnetic spin splitting and valley-selective band inversion separate the two mirror-protected edge modes towards distinct $C_{4z}$-related valleys. We further propose a $\mathrm{PbSe/V_2Se_2O}$ heterobilayer as a feasible platform for realizing this phase through the altermagnetic proximity effect.

\textit{Tight-binding model.}
We consider a square-octagon lattice comprising four sublattices (A--D) per unit cell [Fig.~\ref{fig1}(a)]~\cite{mukherjee2025nontrivial}. The altermagnetic state consists of opposite spin moments on the (A,D) and (B,C) sublattices along the $z$ direction. This collinear spin configuration has zero net magnetization and is invariant under the spin-lattice symmetry $[C_2\parallel C_{4z}]$, where $C_{4z}$ denotes a fourfold lattice rotation and $C_2$ a spin flip operation. For the spin quantization axis along $z$, the magnetic structure also preserves the spin-lattice symmetries $[C_2\parallel \mathcal{M}_{110}]$ and $[C_2\parallel \mathcal{M}_{1\bar{1}0}]$, as well as the mirror symmetry $\mathcal{M}_{001}$. The corresponding tight-binding Hamiltonian is
\begin{equation}
H=H_t+H_{\rm SOC}+H_m,
\label{eq:Hreal}
\end{equation}
where
\begin{align}
H_t&=
-t_1\sum_{\langle i,j\rangle}c_i^\dagger c_j
-t_2\sum_{\langle i,j\rangle'}c_i^\dagger c_j
+{\rm H.c.},\\
H_{\rm SOC}
&=
i\lambda_{\rm SOC}
\sum_{\langle\!\langle i,j\rangle\!\rangle}
c_i^\dagger
(\mathbf{e}_{ij}\!\cdot\!\boldsymbol{\sigma})
c_j,\\
H_m
&=
\sum_{i,\alpha}
\eta_\alpha\delta_m
c_{i\alpha}^\dagger
\sigma_z
c_{i\alpha},
\end{align}
represent the nearest-neighbor hopping, intrinsic Kane-Mele SOC~\cite{kane2005quantum}, and the sublattice-dependent altermagnetic exchange field, respectively. The parameters $t_1$ ($t_2$) denote the intra(inter)-cell hopping amplitudes, $\lambda_{\rm SOC}$ is the SOC strength, and $\eta_{A,D}=-1$ and $\eta_{B,C}=+1$. The exchange field breaks time-reversal symmetry while preserving the spin-lattice symmetries $[C_2\parallel C_{4z}]$, $[C_2\parallel \mathcal{M}_{110}]$, and $[C_2\parallel \mathcal{M}_{1\bar{1}0}]$, and zero net magnetization.

Under periodic boundary conditions,
$
H=\sum_{\mathbf{k}}
\psi_{\mathbf{k}}^\dagger
\hat{\mathcal H}(\mathbf{k})
\psi_{\mathbf{k}},
$
where
$
\psi_{\mathbf{k}}^\dagger=
[\psi_{\mathbf{k}\uparrow}^\dagger,
\psi_{\mathbf{k}\downarrow}^\dagger],
$
with
$
\psi_{\mathbf{k}\sigma}^\dagger=
[c_{A\mathbf{k}\sigma}^\dagger,
c_{B\mathbf{k}\sigma}^\dagger,
c_{C\mathbf{k}\sigma}^\dagger,
c_{D\mathbf{k}\sigma}^\dagger].
$
The Bloch Hamiltonian is
\begin{equation}
\hat{\mathcal H}(\mathbf{k})
=
-I_{2}\otimes\hat{\mathcal H}_{t}(\mathbf{k})
+\sigma_z\otimes\hat{\mathcal H}_{m}(\mathbf{k})
+\sigma_z\otimes\hat{\mathcal H}_{\rm SOC}(\mathbf{k}),
\label{eq:Hbloch}
\end{equation}
where
\begin{align}
\hat{\mathcal H}_{t}(\mathbf{k})&=
\begin{pmatrix}
0&\gamma_1&\gamma_2&\eta_3\\
\gamma_1^{*}&0&\eta_4&\gamma_2\\
\gamma_2^{*}&\eta_4^{*}&0&\gamma_1\\
\eta_3^{*}&\gamma_2^{*}&\gamma_1^{*}&0
\end{pmatrix},
\nonumber\\
\hat{\mathcal H}_{m}(\mathbf{k})&=
\delta_m
\begin{pmatrix}
-1&0&0&0\\
0&1&0&0\\
0&0&1&0\\
0&0&0&-1
\end{pmatrix},
\nonumber\\
\hat{\mathcal H}_{\rm SOC}(\mathbf{k})
&=
2i\lambda_{\rm SOC}
\begin{pmatrix}
0&A&B&0\\
A^{*}&0&0&-B\\
B^{*}&0&0&-A\\
0&-B^{*}&-A^{*}&0
\end{pmatrix},
\end{align}
with
$\gamma_{1,2}=t_1e^{i\mathbf{k}\cdot\mathbf{e}_{1,2}}$,
$\eta_{3,4}=t_2e^{i\mathbf{k}\cdot\mathbf{e}_{3,4}}$,
$A=\cos(\mathbf{k}\!\cdot\!\mathbf{g}_1/2)e^{-i\mathbf{k}\cdot\mathbf{g}_2/2}$,
and
$B=\cos(\mathbf{k}\!\cdot\!\mathbf{g}_3/2)e^{i\mathbf{k}\cdot\mathbf{g}_4/2}$.
The nearest- and next-nearest-neighbor vectors $\mathbf{e}_i$ and $\mathbf{g}_i$ are given in the Supplemental Material (SM)~\cite{supplemental}. Throughout this work, we set $t_1=1$ and express all energies in units of $t_1$.

\textit{Altermagnetic spin splitting and Dirac nodes.}
The electronic structure of the square-octagon lattice without SOC ($\lambda_{\rm SOC}=0$) is shown in Figs.~\ref{fig1}(b)–(d) for $t_2=0.8$ and $\delta_m=0.2$. The spin-resolved Fermi contours in Fig.~\ref{fig1}(b) exhibit the characteristic $d$-wave spin splitting of altermagnets. The spin-up and spin-down Fermi contours are related by a $\pi/2$ rotation through the spin-lattice symmetry $[C_2\parallel C_{4z}]$. The spin splitting vanishes along the mirror-invariant $\Gamma$--$M$ directions, where the mirror symmetries $\mathcal{M}_{110}$ and $\mathcal{M}_{1\bar{1}0}$ interchange the opposite spin-states and enforce their degeneracy. Consequently, the spin polarized states becomes localized at the symmetry-related $X$ and $Y$ valleys, giving rise to a spin-valley-locked electronic structure.

The corresponding band structure in Fig.~\ref{fig1}(c) exhibits four symmetry-related Dirac crossings around the Fermi level along the $\Gamma$--$X(Y)$ and $M$--$X(Y)$ directions. These crossings originate from the momentum-dependent altermagnetic spin splitting. In the absence of altermagnetic order ($\delta_m=0$), the bands remain spin degenerate. A finite exchange field lifts this degeneracy and reverses the ordering of the valence and conduction bands between the $X$ and $Y$ valleys, producing symmetry-related crossings between same-spin bands. Figure~\ref{fig1}(d) shows the Brillouin-zone locations of these Dirac nodes and their evolution with increasing altermagnetic exchange field $\delta_m$. As $\delta_m$ increases, the Dirac nodes move toward the $X(Y)$ valleys. At the critical value $\delta_m=0.8$, pairs of Dirac nodes merge at the valleys to form quadratic band-touching points (see SM~\cite{supplemental}). Further increasing $\delta_m$ annihilates the Dirac nodes and opens a full bulk gap.

The topological nature of these Dirac nodes is characterized by evaluating the winding number $w=\gamma/\pi$, where the Berry phase $\gamma=\oint_{\mathcal C}\mathbf{\mathcal A}(\mathbf{k})\cdot d\mathbf{k}$ is accumulated along a closed loop $\mathcal C$ enclosing the crossing and $\mathbf{\mathcal A}(\mathbf{k})=i\langle u(\mathbf{k})|\nabla_{\mathbf{k}}|u(\mathbf{k})\rangle$ is the Berry connection~\cite{castro2009electronic}. Each Dirac node carries $w=\pm1$. The nodes along the $\Gamma$--$X(Y)$ and $M$--$X(Y)$ directions have opposite winding numbers and annihilate pairwise at the $X(Y)$ valleys to produce the gapped phase at large $\delta_m$.

\begin{figure}[t!]
\centering
\includegraphics[width=0.49\textwidth]{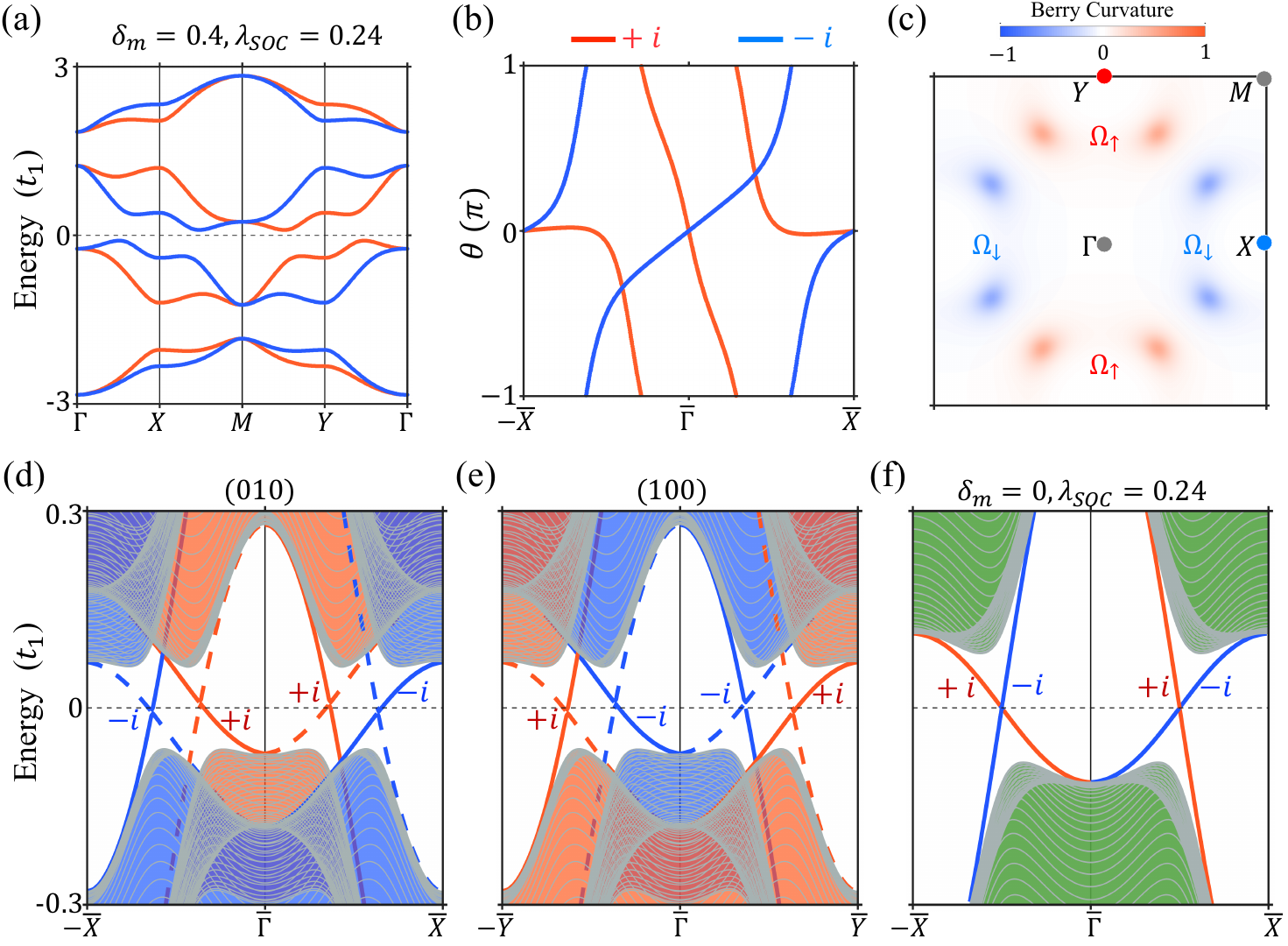}
\caption{(a) Spin-resolved band structure with SOC for $t_2=0.8$, $\lambda_{\rm SOC}=0.24$, and $\delta_m=0.4$. (b) Evolution of the Wannier charge centers in the $\mathcal{M}_{001}=\pm i$ mirror sectors. (c) Berry-curvature distribution in the first Brillouin zone projected onto the $\mathcal{M}_{z}=\pm i$ mirror sectors. (d),(e) Edge spectra of ribbons along the $(010)$ and $(100)$ directions, respectively. Red and blue denote edge modes with mirror eigenvalues $+i$ and $-i$, while solid and dashed curves indicate states localized on opposite edges. (f) Edge spectrum for $\delta_m=0$, corresponding to the conventional mirror Chern insulator with a pair of $\mathcal{T}$-related Dirac cones.}
\label{fig2}
\end{figure}

\textit{Type-II mirror Chern insulator.}
We next examine the effect of intrinsic SOC on these altermagnetic Dirac nodes. As shown in Fig.~\ref{fig2}(a), SOC gaps the Dirac crossings and produces a bulk insulating gap at half filling. The Berry-curvature distribution in Fig.~\ref{fig2}(c) exhibits pronounced positive and negative hotspots around the symmetry-related valleys. Despite broken $\mathcal{T}$ symmetry, the Berry-curvature contributions from the $C_{4z}$-related valleys cancel under the spin-lattice symmetry $[C_2\parallel C_{4z}]$, giving a zero Chern number. Since the system preserves the mirror symmetry $\mathcal{M}_{001}$, the Hamiltonian decomposes into two mirror sectors with eigenvalues $\pm i$. This allows calculations of the mirror Chern number,
$
C_{\mathcal M}=\frac{C_{+i}-C_{-i}}{2},
$
where $C_{+i}$ and $C_{-i}$ denote the Chern numbers of the $+i$ and $-i$ mirror sectors, respectively~\cite{hsieh2012topological,antonenko2025mirror}. The Wilson-loop evolution of the Wannier charge centers in Fig.~\ref{fig2}(b) yields $C_{+i}=2$ and $C_{-i}=-2$. This results in a mirror Chern number of $C_{\mathcal M}=2$.

This nontrivial bulk topology is further confirmed by the edge spectra of the $(010)$ and $(100)$ ribbons shown in Figs.~\ref{fig2}(d) and \ref{fig2}(e), where the solid and dashed lines denote states localized on opposite edges. In both ribbon geometries, two pairs of counterpropagating edge states connect the valence and conduction bands, consistent with the calculated mirror Chern number $C_{\mathcal M}=2$. Since the system preserves the mirror symmetry $\mathcal{M}_{001}$ ($\mathcal{M}_{001}^{2}=-1$), the edge states remain eigenstates of the mirror operator with eigenvalues $\pm i$. Consequently, the crossings between opposite mirror eigenstates are symmetry protected.

\begin{figure}[b!]
\centering
\includegraphics[width=0.49\textwidth]{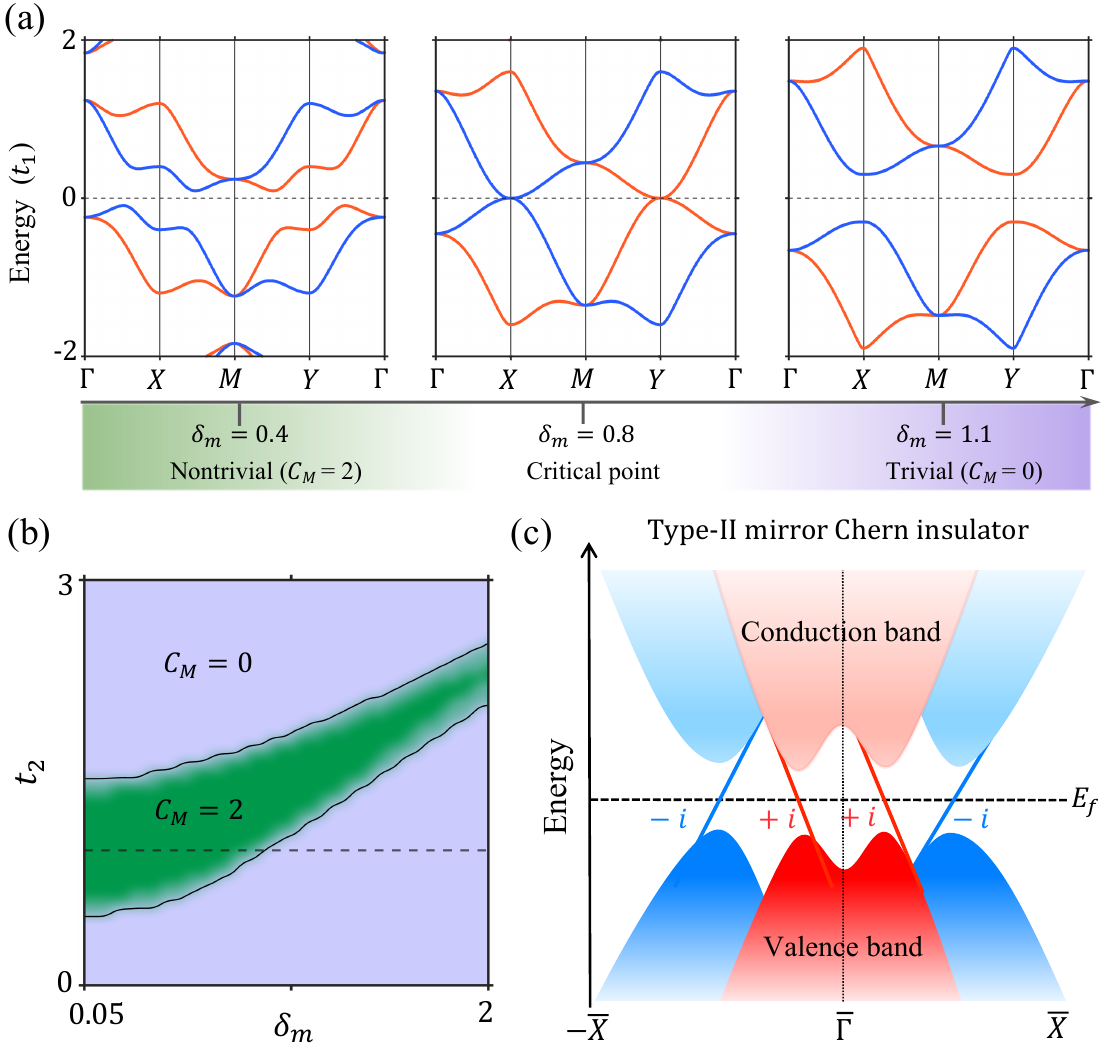}
\caption{(a) Bulk band structure with SOC as a function of the altermagnetic exchange field $\delta_m$. (b) Topological phase diagram in the $t_2$--$\delta_m$ parameter space for $\lambda_{\rm SOC}=0.24$. The color scale denotes the mirror Chern number $C_{\mathcal M}$. (c) Schematic illustration of the type-II mirror Chern insulator. Counterpropagating edge modes with opposite mirror eigenvalues are separated into symmetry-related valleys of the Brillouin zone.}
\label{fig3}
\end{figure}

Unlike conventional mirror Chern insulators, where the two edge modes with opposite mirror eigenvalues form a Dirac cone pinned at the band-inversion momentum, altermagnetic spin splitting reconstructs the edge-state connectivity. For the $(010)$ ribbon, along the $\bar{\Gamma}$--$\bar{X}$ direction, the backward-moving $+i$ branch connects the spin-up conduction states at $\bar{\Gamma}$ (projected bulk $Y$ valley), whereas the forward-moving $-i$ branch connects the spin-down conduction states at the $\bar{X}$ valley. Their crossing shifts below the Fermi level and forms an electron-like edge pocket. Along the $-\bar{X}$--$\bar{\Gamma}$ direction, the edge-mode connectivity is reversed, shifting the crossing above the Fermi level to form a hole-like edge pocket. The two crossings are therefore separated in both energy and momentum, rather than forming Dirac cones pinned at the Fermi level. For the $(100)$ ribbon [Fig.~\ref{fig2}(e)], the branch connectivity and spin polarization are interchanged by the underlying $C_{4z}$ spin-lattice symmetry.

For comparison, Fig.~\ref{fig2}(f) shows the $(010)$ edge spectrum in the absence of the altermagnetic exchange field ($\delta_m=0$). The two edge modes with opposite mirror eigenvalues reconnect symmetrically to the inverted bulk bands and cross at the Fermi level along the $\bar{\Gamma}$--$\bar{X}$ direction to form a mirror-protected Dirac cone. Along the $-\bar{X}$--$\bar{\Gamma}$ direction, the branch velocities are reversed by $\mathcal{T}$ symmetry, giving the time-reversal-partner Dirac cone. The pair of $\mathcal{T}$-related Dirac cones therefore resides on the mirror-invariant line, recovering the conventional mirror Chern insulator. In contrast, altermagnetic spin splitting preserves the mirror eigenvalues but reconstructs the edge-state connectivity, separating the mirror-protected crossings in momentum and energy. This momentum-separated edge spectrum constitutes the defining boundary signature of the type-II mirror Chern insulator.

\textit{Phase diagram and tunability.}
Figure~\ref{fig3}(a) shows the evolution of the bulk band structure with the altermagnetic exchange field $\delta_m$ for a fixed SOC strength $\lambda_{\rm SOC}=0.24$. Starting from the mirror Chern insulating phase, increasing $\delta_m$ continuously reduces the bulk gap until it closes at the critical value $\delta_m=0.8$. Further increasing $\delta_m$ reopens the gap and removes the band inversion, driving a topological phase transition from the mirror Chern insulator with $C_{\mathcal M}=2$ to a trivial insulator with $C_{\mathcal M}=0$. The corresponding phase diagram in the $(t_2,\delta_m)$ parameter space is shown in Fig.~\ref{fig3}(b). The mirror Chern insulator phase occupies a broad region of the phase diagram and is separated from the trivial phase by bulk-gap-closing boundaries. The topology can therefore be tuned by varying either the intercell hopping $t_2$ or the altermagnetic exchange field $\delta_m$.

Figure~\ref{fig3}(c) schematically illustrates the edge-state connectivity of the type-II mirror Chern insulator. Two pairs of counterpropagating edge modes associated with the mirror sectors $\mathcal{M}_{001}=\pm i$ traverse the bulk gap. Unlike conventional mirror Chern insulators, where the two edge modes with opposite mirror eigenvalues form a Dirac cone at the same momentum, altermagnetic spin splitting separates the edge modes into symmetry-related valleys but preserves their mirror eigenvalues. This momentum-separated edge-state connectivity distinguishes the type-II mirror Chern insulator from conventional mirror Chern insulators.

\begin{figure}[t!]
\centering
\includegraphics[width=0.49\textwidth]{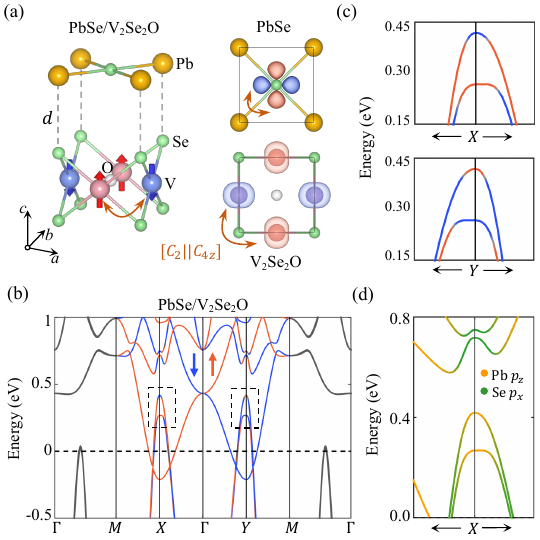}
\caption{(a) Crystal structure of the PbSe/$\mathrm{V_2Se_2O}$ heterobilayer. The two magnetic sublattices of $\mathrm{V_2Se_2O}$ are related by the spin-lattice symmetry $[C_{2}\parallel C_{4z}]$. The right panels show the real-space spin density centered on a Se atom in PbSe (top) and a V atom in $\mathrm{V_2Se_2O}$ (bottom). (b) Spin-resolved band structure of the PbSe/$\mathrm{V_2Se_2O}$ heterobilayer for an interlayer separation of $d=3.5$~\AA\ with SOC. Red and blue denote spin-up and spin-down $S_z$ polarization, respectively. (c) Enlarged view of the PbSe-derived bands near the $X$ and $Y$ valleys. (d) Orbital-resolved band structure of the PbSe-derived states near the $X$ valley. Orange and green denote the Pb-$p_z$ and Se-$p_x$ orbital contributions, respectively.}
\label{fig4}
\end{figure}

\textit{Possible material realization.}
Having established the type-II mirror Chern insulator in the model, we now consider its realization in a realistic material. A promising route is to combine a 2D mirror Chern insulator with an altermagnet through the proximity effect~\cite{AM_proximity2026}. As a representative example, we investigate a PbSe/$\mathrm{V_2Se_2O}$ heterobilayer using first-principles density functional theory (see SM for methods)~\cite{supplemental}. Monolayer PbSe is a two-dimensional mirror Chern insulator with $C_{\mathcal M}=2$~\cite{Wrasse2014}, whereas monolayer $\mathrm{V_2Se_2O}$ is a $d$-wave altermagnet with valley-selective spin splitting at the symmetry-related $X$ and $Y$ valleys~\cite{jiang2025metallic,zhang2025crystal,Zhan2025multiphysics}. To examine the weak-coupling proximity effect while preserving the mirror-topological phase of PbSe, we set the interlayer separation to $d=3.5$~\AA.

Figure~\ref{fig4}(a) shows the real-space spin density of the PbSe/$\mathrm{V_2Se_2O}$ heterobilayer. Besides the intrinsic $d$-wave spin density of $\mathrm{V_2Se_2O}$, a finite altermagnetic spin polarization appears in the initially nonmagnetic PbSe layer. This demonstrates the transfer of the altermagnetic exchange field across the interface and confirms that PbSe inherits the altermagnetic character of $\mathrm{V_2Se_2O}$. The spin-resolved band structure with SOC in Figs.~\ref{fig4}(b)-\ref{fig4}(c) shows that the $\mathrm{V_2Se_2O}$ bands remain weakly hybridized with the PbSe states, whereas the PbSe-derived bands exhibit characteristic valley-selective spin splitting. The band inversion between the Pb-$p_z$ and Se-$p_{x(y)}$ states remains intact [Fig.~\ref{fig4}(d)], preserving the mirror Chern number $C_{\mathcal M}=2$. More importantly, the PbSe-derived states exhibit locally asymmetric spin splitting around the $X$ and $Y$ valleys. This valley spin asymmetry is expected to separate the mirror-protected edge modes and realize the proposed type-II mirror Chern insulator. Beyond the proximity effect, the type-II mirror Chern insulator may also arise intrinsically. Promising candidates include strained $\mathrm{KRu_4O_8}$ with a square-octagon lattice~\cite{smejkal2022beyond} and bottom-up predicted square-lattice materials~\cite{Verma2026}, which have the potential to host altermagnetic order. These materials exhibit valley-polarized electronic states at the $X$ and $Y$ valleys with small band gaps and may therefore host the proposed type-II mirror Chern insulator under suitable conditions.

\textit{Summary.} We predict a new type of mirror Chern insulator in two-dimensional altermagnets, termed a type-II mirror Chern insulator, where altermagnetic spin splitting reconstructs the conventional mirror-protected edge spectrum into momentum-separated edge modes. Using a square-octagon lattice model, we show that valley-polarized Dirac nodes generated by the $[C_{2}\parallel C_{4z}]$ spin-lattice symmetry are gapped by SOC and form a mirror Chern insulator with $C_{\mathcal M}=2$. Unlike conventional mirror Chern insulators, altermagnetic spin splitting separates the edge modes into symmetry-related valleys and gives rise to the characteristic type-II boundary spectrum. We further propose the PbSe/$\mathrm{V_2Se_2O}$ heterobilayer as a feasible material platform for realizing this phase. More broadly, our work establishes altermagnetism as a route to crystalline topological phases with unconventional boundary-state connectivity and momentum-separated edge modes.

\section{Acknowledgement}
This work is supported by the Department of Atomic Energy, Government of India, under Project Identification Nos. RTI4013 and RTI4015.

\bibliography{ref}
\end{document}